\documentclass[pre,aps,showpacs,superscriptaddress]{revtex4}
\usepackage{graphicx}
\begin{document}

\title{Lyapunov Instability versus Relaxation Time in two Coupled Oscillators}

\author{P.K. Papachristou}
\affiliation{Department of Physics, University of Athens, GR--15771,
Athens,
Greece}

\author{E. Mavrommatis}
\affiliation{Department of Physics, University of Athens, GR--15771,
Athens,
Greece}

\author{V. Constantoudis}
\affiliation{Institute of Microelectronics (IMEL), NCSR "Demokritos",
P.O. Box 60228,
Aghia Paraskevi, Attiki, Greece 15310 and Physics Department, National
Technical University,
Athens, Greece}

\author{F.K. Diakonos}
\affiliation{Department of Physics, University of Athens, GR--15771,
Athens,
Greece}

\author{J. Wambach}
\affiliation{Institut f\"ur Kernphysik, Technische Universit\"at Darmstadt, Schlossgartenstr. 9, D-64289 Darmstadt, Germany}

\date{\today}

\begin{abstract}
We consider the relation between relaxation time and the largest Lyapunov exponent in a system of two coupled oscillators, one of them being harmonic. It has been found that in a rather broad region of parameter space, contrary to the common expectation, both Lyapunov exponent and relaxation time increase as a function of the total energy. This effect is attributed to the fact that above a critical value of the total energy, although the Lyapunov exponent increases, KAM tori appear and the chaotic fraction of phase space decreases. We examine the required conditions and demonstrate the key role of the dispersion relation for this behaviour to occur. This study is useful, among others, in the understanding of the damping of nuclear giant resonances.
\end{abstract}


\maketitle

\section{Introduction}
It is well known that time-independent Hamiltonian systems with at least two degrees of freedom can exhibit chaotic behaviour \cite{LIC83,TAB89}. In most cases the dynamics is mixed, i.e. regular and chaotic regions coexist in phase space. In general, the largest Lyapunov exponent ($\lambda_1$) is the most common measure of the chaoticity of the system. It appears that, at least in completely chaotic systems, increasing $\lambda_1$, the relaxation time, defined as the time required for an observable to reach its equilibrium value, decreases \cite{KAN94,KANa04}. However, in mixed systems the relationship between $\lambda_1$ and relaxation time is not well established yet.

Relaxation time is an important tool in many body problems \cite{KAN89,KAN93,BER94,BUR95,KAN97,TSU96,BAL98,LEP98,TSU00,KAN03,KAN04}. Let us suppose that a many body system is collectively excited by an external perturbation which for example may be a collision with a particle-projectile or a laser pulse. The energy of the collective excitation is gradually absorbed by the internal dynamics of the system and the time needed for this absorption to occur is the relaxation time of the excitation.

Such relaxation processes can be encountered for example in the field of nuclear physics and in particular in the damping of nuclear giant resonances \cite{HAR04,NEU01,BLA80}. In the latter the system has mixed dynamics. The mechanism responsible for the dissipation of nuclear collective energy is still an open problem. Especially, the effects of chaotic dynamics on the damping of nuclear excitations is still not completely understood. One body dissipation \cite{BLO78} refers to processes involving a single nucleon (with mean free path of the order of the size of the nucleus) interacting with the collective nuclear potential. We have employed a simple classical model based on this independent particle approach for the dynamics of the nucleons. This model can be used for the interpretation of the isoscalar monopole resonance \cite{YOU99}(the breathing mode) decay \cite{BUR95,BAL98,DRO95,VRE99}. It consists of a harmonic oscillator describing the collective excitation (to a first approximation) coupled with a nonlinear (Woods-Saxon) oscillator, which, in the independent particle approach, represents the motion of each nucleon in the average potential created by all nucleons. The model corresponds to a time-independent Hamiltonian system with 2 degrees of freedom. The detailed investigation of the specific giant resonances through this simple classical model will be given elsewhere \cite{PAP05}. Here, we will use this model in order to elucidate the relationship of relaxation dynamics with the structure of the chaotic component of the phase space of the system. We should add that an appropriate classical model has also been used to study the damping of giant dipole resonance in Ref. \cite{PAP00}. While the intensity of that chaotic component is usually measured by the largest Lyapunov exponent as mentioned above, its extent is quantified by the fraction of phase-space occupied by chaotic orbits. The aim of this paper is to explore the interrelation between the relaxation time of a harmonic oscillator and $\lambda_1$, as well as how this is affected by changes in the chaotic fraction of phase space.

More specifically, as the considered model posseses mixed phase space, we have found that, contrary to a completely chaotic system, a broad range of parameter space exists where $\lambda_1$ increases as the total energy increases and also both $\lambda_1$ and the relaxation time of an ensemble of chaotic orbits increase. We demostrate that this relationship is affected by the form of the dispersion relation $\omega(E)$ of the nonlinear oscillator. In particular, the frequency $\omega(E)$ must either be a strictly increasing function of the energy or have a maximum (zero dispersion oscillator \cite{SOS03}) for the above mentioned behaviour to occur. In these cases, the chaotic fraction of phase space is shrinking as the total energy increases, inducing a delay in the relaxation of the dynamics. Finally, the dependence of the results on the form of the coupling as well as on the particular choice of the potential of the non-harmonic oscillator is explored.

The paper consists of four further sections. In Sec. II we describe our model system. In Sec. III we present our findings on the relation between relaxation time and Lyapunov exponent for a particular case. In Sec. IV we discuss the dependence on the system parameters of the effects presented in Sec. III. In Sec. V we summarize our main results and give future prospects. 

\section{Description of the model system}
Our model system consists of a particle moving in a Woods-Saxon well of finite depth $V_o$ coupled with a harmonic oscillator moving with frequency $\omega_{HO}$. The Hamiltonian of the system is
\begin{equation}
H = \frac{{p_r^2 }}{{2m}} + \frac{{p_R^2 }}{{2M}} + V(r,R), 
\end{equation}
where the potential $V(r,R)$ is given by
\begin{equation}
V(r,R)=- \frac{{V_o }}{{1 + \exp \left( {{\textstyle{{r - R_o  + b(R_0  - R)} \over a}}} \right)}} + \frac{1}{2}M\omega_{0} ^2 (R - R_0 )^2 .
\end{equation}
A contour plot of the potential for a specific choice of the parameters, namely $a=0.5 fm$, $b=0.2$, $R_0=7.69 fm$, $V_0=48  MeV$, $M/m=1$ and $\omega_{0}=13.73 MeV/\hbar$, is shown in Fig.1. 

The choice of this Hamiltonian is motivated by the damping of a density collective mode in nuclei. Specifically, this mode is represented by a harmonic oscillator of mass $M$ which interacts with the nucleons, represented as independent particles of mass $m$ moving in an average potential which has the form of a Woods-Saxon well. The strength of the coupling between the particle and the harmonic oscillator is controlled by the parameter $b$. In the limit $b\to 0$ the oscillators become uncoupled and thus the system becomes integrable. As $b$ increases the system becomes increasingly chaotic. Hamilton's equations of motion for the system are
\begin{equation}
\mathop r\limits^.  = \frac{{p_r }}{m},
\end{equation}
\begin{equation}
\mathop {p_r }\limits^.  = - \frac{{V_0 }}{a}\frac{{\exp \left( {{\textstyle{{r - R_o  + b(R_0  - R)} \over a}}} \right)}}{{\left( {1 + \exp \left( {{\textstyle{{r - R_o  + b(R_0  - R)} \over a}}} \right)} \right)^2 }},
\end{equation}
\begin{equation}
\mathop R\limits^.  = \frac{{p_R }}{M},
\end{equation}
\begin{equation}
\mathop {p_R }\limits^.  =   \frac{{bV_0 }}{a}\frac{{\exp \left( {{\textstyle{{r - R_o  + b(R_0  - R)} \over a}}} \right)}}{{\left( {1 + \exp \left( {{\textstyle{{r - R_o  + b(R_0  - R)} \over a}}} \right)} \right)^2 }} - M\omega_{0}^2 \left( {R - R_0 } \right).
\end{equation}
These equations are solved with the use of an adaptive fourth order Runge-Kutta algorithm. The total energy $E$ was conserved with relative error $\Delta E/E\leq 10^{-5}$. Rescaling the differential equations of motion we find that the relevant parameters of the system are $M/m$, $R_0/a$, $V_0/(ma^2\omega_{0}^2)$ and $b$.
\begin{figure}
\begin{center}
\includegraphics[height=8cm]{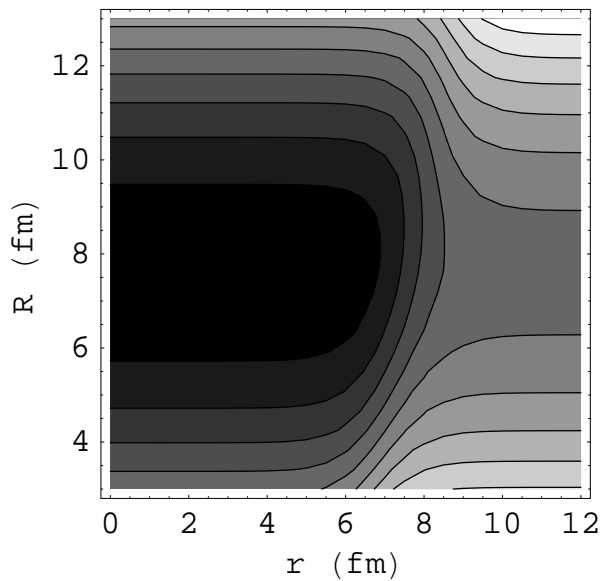}
\caption{Contour plot of the potential $V(r,R)$ of the system for $a=0.5 fm$, $b=0.2$, $R_0=7.6 fm$, $V_0=48 MeV$ and $\omega_{0}=13.73 MeV/\hbar$ (see Eq. 2).}
\end{center}
\end{figure}

\section{Relaxation time, Lyapunov exponent and chaotic fraction of phase space}
For the analysis of the system dynamics, we start with a microcanonical ensemble of initial conditions (5000 in most of the cases) in a rectangular region of $(r,E_p)$ plane, where $E_p$ is the particle energy defined as
\begin{equation}
E_p  = \frac{{p_r^2 }}{{2m}} - \frac{{V_o }}{{1 + \exp \left( {{\textstyle{{r - R_o  + b(R_0  - R)} \over a}}} \right)}}.
\end{equation}
For each such ensemble, the total energy $E=E_p+E_{HO}$ is constant, where $E_{HO}$ is the energy of the harmonic oscillator given by
\begin{equation}
E_{HO}={{1 \over 2}}M\omega_{0} ^2 (R - R_0 )^2 + {p_R^2 \over {2M}}.
\end{equation}
All initial conditions we use belong to the chaotic region of phase space. We evolve them forward in time and at each time instant $t$ we calculate the mean value $<R(t)>$ of $R$. It is expected that $<R(t)>$ will approach the equilibrium value $R_{eq}$ (which is very close to $R_0$) almost exponentially \cite{KANa04}, since all orbits in the ensemble are chaotic. The corresponding time scale defines the relaxation time. For the values of the parameters chosen, the time dependence of $<R(t)>-R_{eq}$  can be fitted very well by a function of the form $R - R_{eq}  = Ae^{ - \gamma t} \cos \left({\omega t}\right)$. We calculate the half-life $T_{1/2}$ of the initial amplitude $A$ as a function of the total energy of the system. $T_{1/2}$ and $\gamma$ are related by $T_{1/2}=\ln 2/\gamma$. The structure of the phase space allows us, as it will become clear later, to use the same rectangular region of initial conditions for all the considered values of the total energy. However, we have found that there is no dependence of the relaxation time on the location of the distribution of the initial conditions in phase space, provided that this distribution lies entirely in the chaotic regime. 

To quantify the degree of chaoticity of these orbits, we calculate for each one the Lyapunov exponent $\lambda_1$ by averaging the growth rates of small perturbations along them \cite{LIC83,TAB89}. It is found numerically that, for sufficiently large integration times (of the order of the relaxation time), the average value $<\lambda_1>$ calculated in this way is independent of  the location of the rectangle of the initial conditions in the chaotic region. We determine the average value $<\lambda_1>$ over the ensemble of initial conditions as a function of the total energy of the system.

In addition to $\lambda_1$ we calculate the percentage q of chaotic phase space in the $(R,p_R)$ plane as a function of the total energy of the system \cite{CON87,VAR00}. In order to do this, we start with an ensemble of a large number of chaotic initial conditions on the $(R,p_R)$ section and we integrate them forward in time until they cover the entire accessible chaotic region. We then divide the $(R,p_R)$ section in $N=10^4$ cells and count the number of occupied cells $N_{occ}$, i.e. the number of cells in which there is at least one orbit point. We estimate the percentage of chaotic phase space as $q = {\textstyle{{N_{occ} } \over N}}100\%$.
\begin{figure}
\begin{center}
\includegraphics[width=7cm]{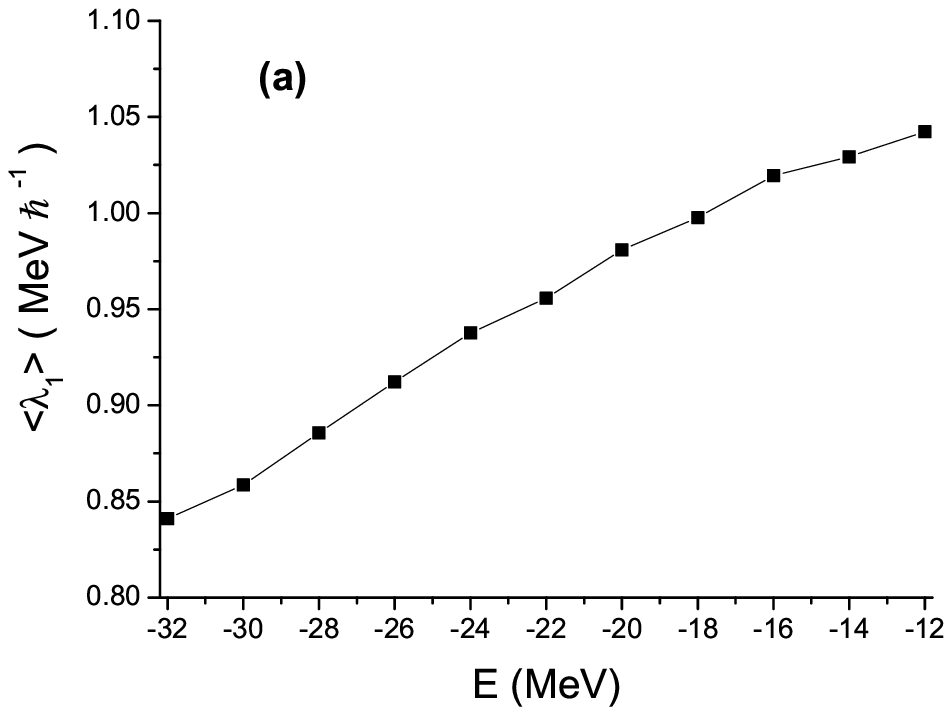}%
\includegraphics[width=7cm]{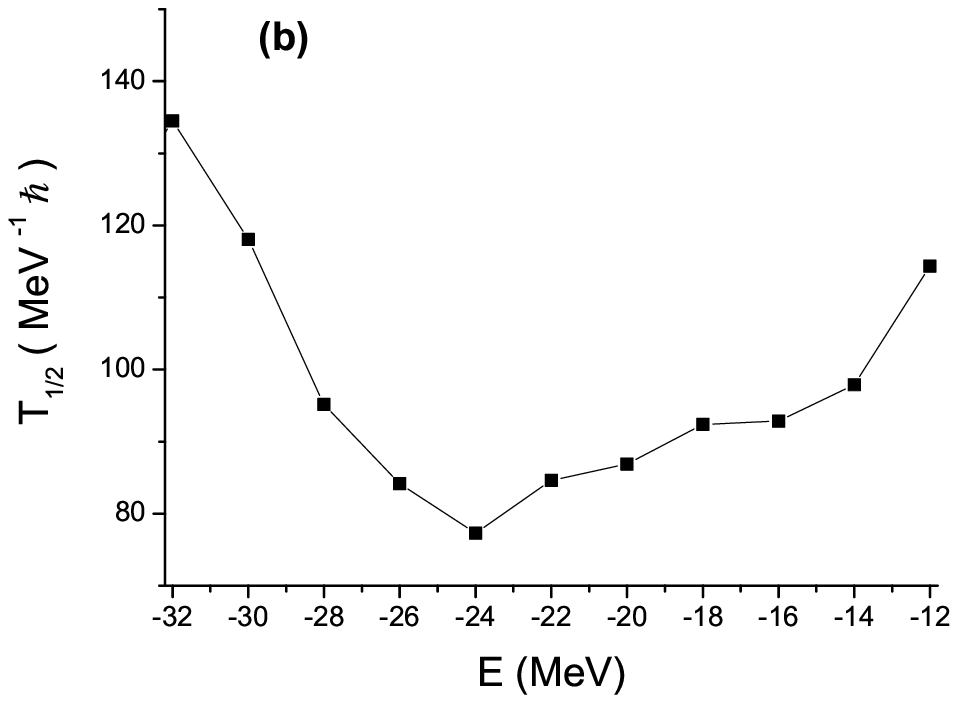}
\includegraphics[width=7cm]{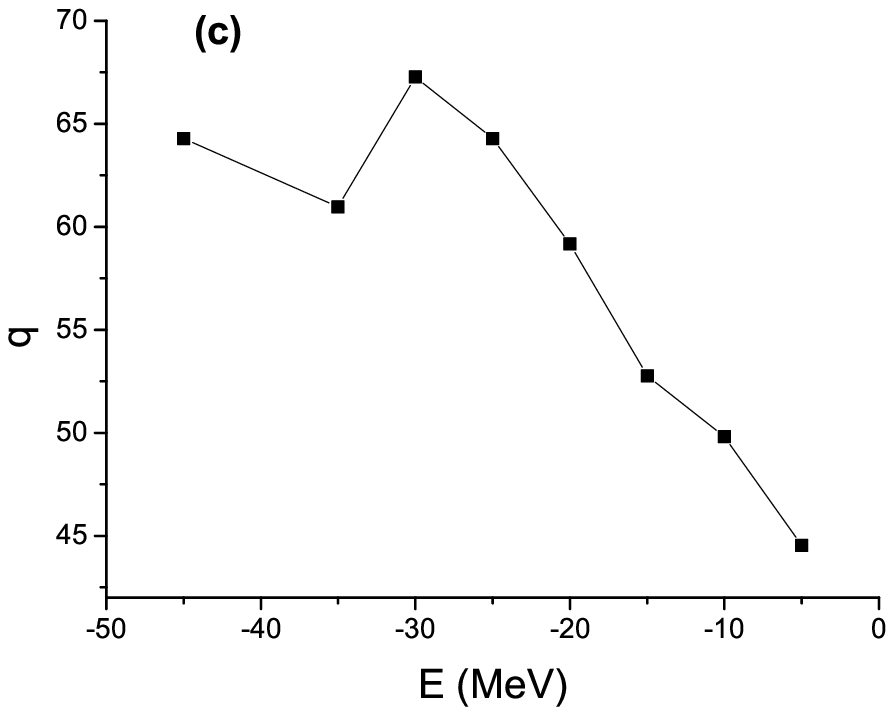}
\caption{(a) $<\lambda_1>$, (b) $T_{1/2}$ and (c) $q$ as a function of the total energy $E$ for $M/m=1$, $R_0=7.69 fm$, $a=0.05 fm$, $V_0=48 MeV$, $\omega_{0}=13.73 MeV/\hbar$ and $b=0.08$.}
\end{center}
\end{figure}

In the study presented in this section we choose the following values for the parameters of the system: $R_0=7.69 fm$, $V_0=48 MeV$ and $\omega_{0}=13.73 MeV/\hbar$. These parameters are relevant for the decay of the isoscalar giant monopole resonance of the nucleus $^{208}$~Pb \cite{YOU99,BUR95,VRE99,BRA88}. The strength of the coupling has been chosen equal to $b=0.08$. We also choose $M/m=1$ and $a=0.05 fm$. The results for $<\lambda_1>$, $T_{1/2}$, and $q$ as a function of the total energy of the system are shown in Fig. 2. It is clearly seen that although the value of $<\lambda_1>$ is a monotonically increasing function of the total energy E (see for instance \cite{KAN94}), the relaxation time, above a critical value of the total energy, stops to decrease and starts increasing contrary to the common expectation from fully chaotic systems. We also observe that above this critical energy, the percentage $q$ of chaoticity in the $(R,p_R)$ section is decreasing. Motivated by this observation we consider in more detail the Poincar\'e surfaces of section for different values of the total energy of the system. In Fig. 3, $(R,p_R)$ sections (for $r=R_0/2$ and $p_r>0$) for four values of the total energy ($E=-35 MeV$, $-25 MeV$, $-15 MeV$ and $-5 MeV$) are shown. In Fig. 4 we plot $(r,E_p)$ sections (for $R=R_0$, $p_R>0$) for the same values of the total energy.
\begin{figure}
\begin{center}
\caption{$(R,p_R)$ Poincar\'e sections for (a) $E=-35 MeV$, (b) $E=-25 MeV$, (c) $E=-15 MeV$ and (d) $E=-5 MeV$. The other parameters are the same as those of Fig. 1.}
\end{center}
\end{figure}
\begin{figure}
\begin{center}
\caption{$(r,E_p)$ Poincar\'e sections for (a) $E=-35 MeV$, (b) $E=-25 MeV$, (c) $E=-15 MeV$ and (d) $E=-5 MeV$. The other parameters are the same as those of Fig. 1.}
\end{center}
\end{figure}

The regular region of Figs. 3(c) and (d) contains invariant tori and corresponds to the upper invariant curves of Figs. 4(c) and (d) respectively. These curves are close to straight lines and correspond to almost constant particle energy and hence almost constant total energy. They are therefore Kolmogorov- Arnold- Moser (KAM) tori \cite{LIC83}, which are remnants of the uncoupled integrable system. By inspecting these Poincar\'e sections, we observe that the appearance of KAM tori leads to the observed relative decrease of chaos in phase space.  The onset of this decrease coincides with a change in the monotonicity of $T_{1/2}$ as a function of the energy. Once the KAM tori appear, they are not destroyed as the energy increases. We conjecture that such an appearance of KAM tori counteracts the increase of the Lyapunov exponent and leads to a change in the monotonicity of the relaxation time. But what causes this reappearance of KAM tori in the system phase space for such large values of the energy? How is it connected to the basic properties of the non-harmonic oscillator ? To answer this question, we investigate the form of the dispersion relation of the non-harmonic oscillator. This relation, which is sensitive to the parameters of the system and controls the appearance of KAM tori, will be studied in the following section.

\section{The key role of dispersion relation and the influence of the coupling strength}

The diffuseness parameter $a$ is of crucial importance for the appearance of the effect described in the previous section. 
In the Woods-Saxon potential $V_{WS} (r) = -V_0 /\left( {1 + \exp \left( {{\textstyle{{r - R_0 } \over a}}} \right)} \right)$ the frequency of the motion of the particle $\omega_p$ depends on its energy $E_p$, whereas the frequency of the harmonic oscillator $\omega_{HO}$ is independent of the energy. The frequency of the motion of a particle in a potential $V(r)$ is given by
\begin{equation}
\omega_p (E_p ) = \pi \left( {\int\limits_{r_l }^{r_r } {\frac{{dr}}{{\sqrt {2\left( {E_p  - V(r)} \right)} }}} } \right)^{ - 1},
\end{equation}
where $r_r$ and $r_l$ are the right and left turning points at a given energy $E_p$, i.e. the roots of the equation $E_p=V(r)$. The form of the $\omega_p(E_p)$ curve of the Woods-Saxon potential depends strongly on $a$ as can be seen from Fig. 5, where $V(r)$ and $\omega_p(E_p)$ curves for four values of $a$ are shown. At the limit $a\to 0$ the potential tends to the square well and therefore the $\omega_p(E_p)$ curve has the form $\omega_p\propto \sqrt{E_p}$. For large values of $a$, $\omega_p(E_p)$ is decreasing and for intermediate values it has a maximum.
\begin{figure}
\begin{center}
\includegraphics[width=7cm]{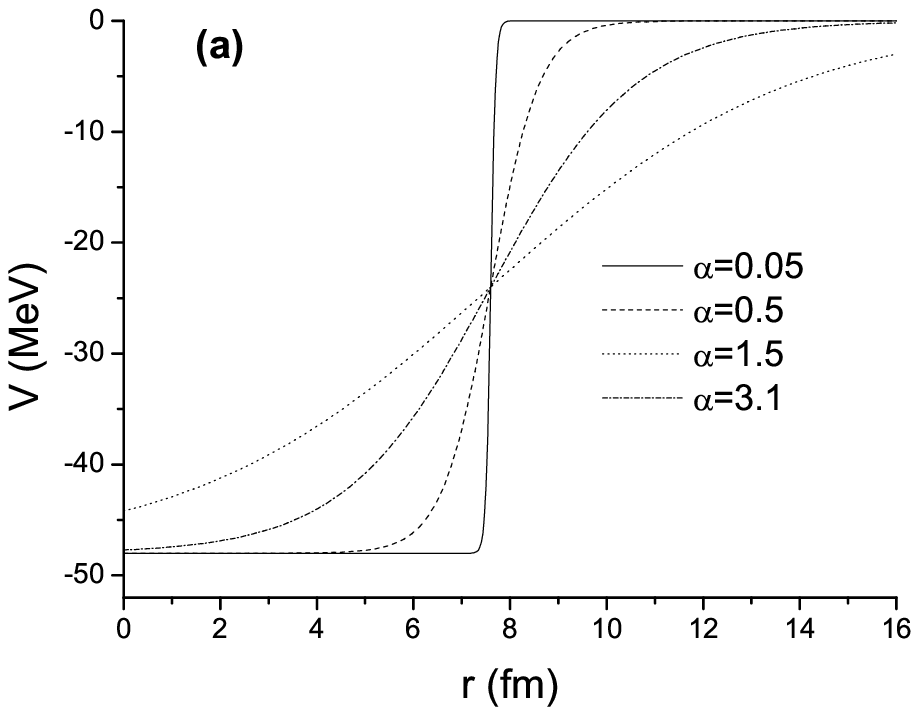}%
\includegraphics[width=7cm]{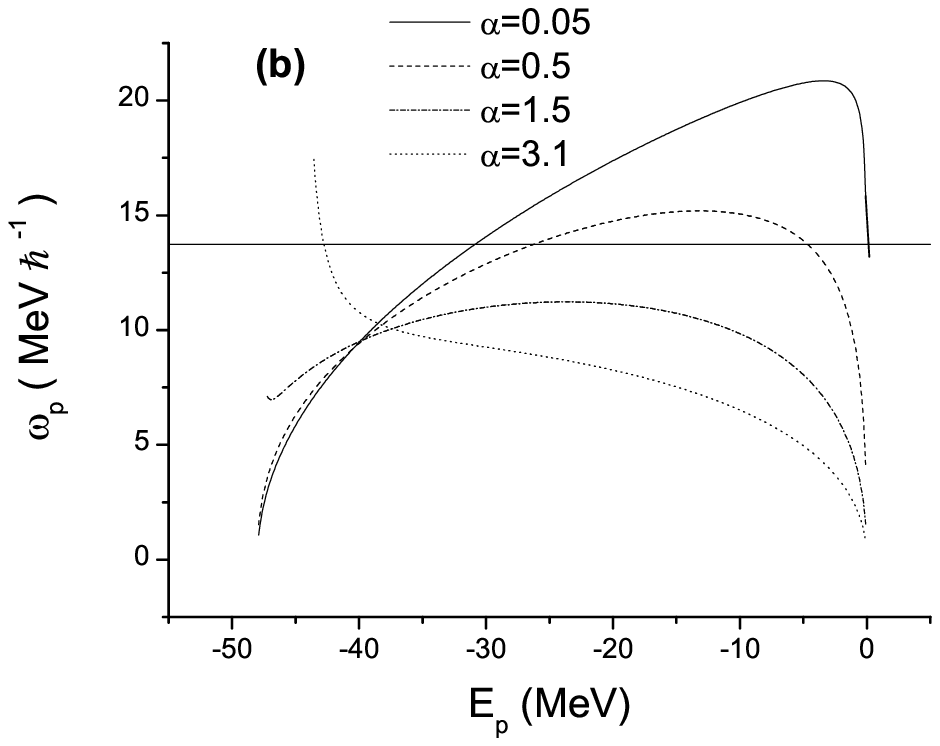}
\caption{(a) The Woods-Saxon potential for $V_0=48 MeV$ and four values of $a$, namely $0.05 fm$, $0.5 fm$, $1.5 fm$, and $3.1 fm$, (b) The $\omega_p(E_p)$ curve of the Woods-Saxon potential for the same values of the $a$. The straight line shows the frequency of the harmonic oscillator.}
\end{center}
\end{figure}

For a fixed value of $a$ (i.e. $a=0.05fm$) for which the observed behaviour of relaxation time occurs, the $\omega_p(E_p)$ curve has two intersection points with the corresponding curve of the harmonic oscillator. The leftmost intersection point is a $\omega_p:\omega_{HO}=1:1$ resonance. This resonance is apparent in Figs. 3 and 4. The rightmost intersection point is a much less apparent $1:1$ resonance located close to the edge of the potential well. For small values of $b$, the phase-space is dominated by KAM tori. Resonances of small width also exist. As the coupling strength $b$ increases, the width of these resonances also increases. They therefore overlap with higher-order resonances located close to them. This overlap leads to the appearance of a chaotic region. The area of this chaotic region increases with $b$. Higher-order resonances ($1:2$, $1:3$, $\ldots$) appear in the low-energy region and their density increases as the energy becomes smaller. Therefore, as $b$ increases {\it chaos emerges first in the low-energy region of the $(r,E_p)$ section}. This transition is illustrated in Fig. 6, where $(r,E_p)$ sections for $a=0.05 fm$ and several values of $b$ are shown.
\begin{figure}
\begin{center}
\caption{$(r,E_p)$ Poincar\'e sections for $E=-5 MeV$, $a=0.05 fm$ and (a) $b=0.02$, (b) $b=0.04$, (c) $b=0.1$, (d) $b=0.15$, (e) $b=0.2$ and (f) $b=0.3$. The other parameters are the same as those of Fig. 1.}
\end{center}
\end{figure}

For fixed values of $b$ and $a$, above some particle energy threshold only KAM tori exist. As the energy increases, the existing KAM tori are not destroyed and new KAM tori are added to the phase space in the region of large particle energies. This happens provided the coupling is not strongly chaotic. Therefore, as the energy increases, the relative area of the chaotic phase space decreases. This can be seen from Fig. 4. Increasing $b$ makes the system more chaotic, i.e. both the Lyapunov exponent and the relative area of the chaotic phase space increase (see Fig. 6). The inverse behaviour of relaxation time and Lyapunov exponent appears as long as a sufficiently wide layer of KAM tori exists at high energies. This can be seen from Fig. 7, where the Lyapunov exponent increases, the percentage of relative chaotic area decreases and the relaxation time, above a critical value of the total energy, either increases or remains almost constant.
\begin{figure}
\begin{center}
\includegraphics[width=7cm]{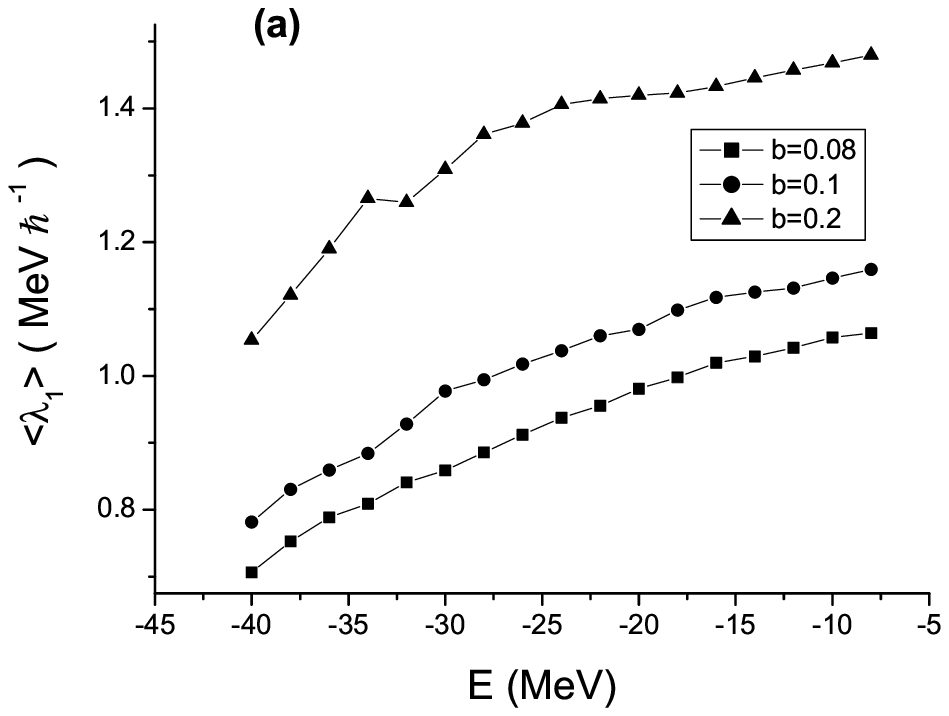}%
\includegraphics[width=7cm]{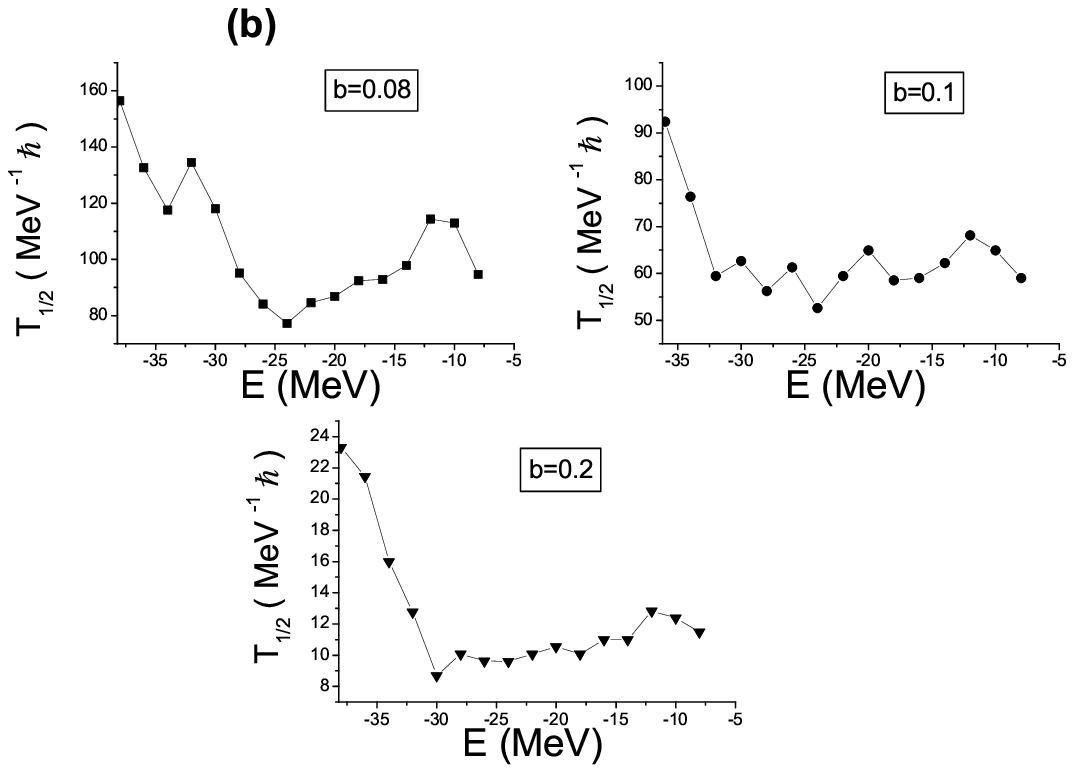}
\includegraphics[width=7cm]{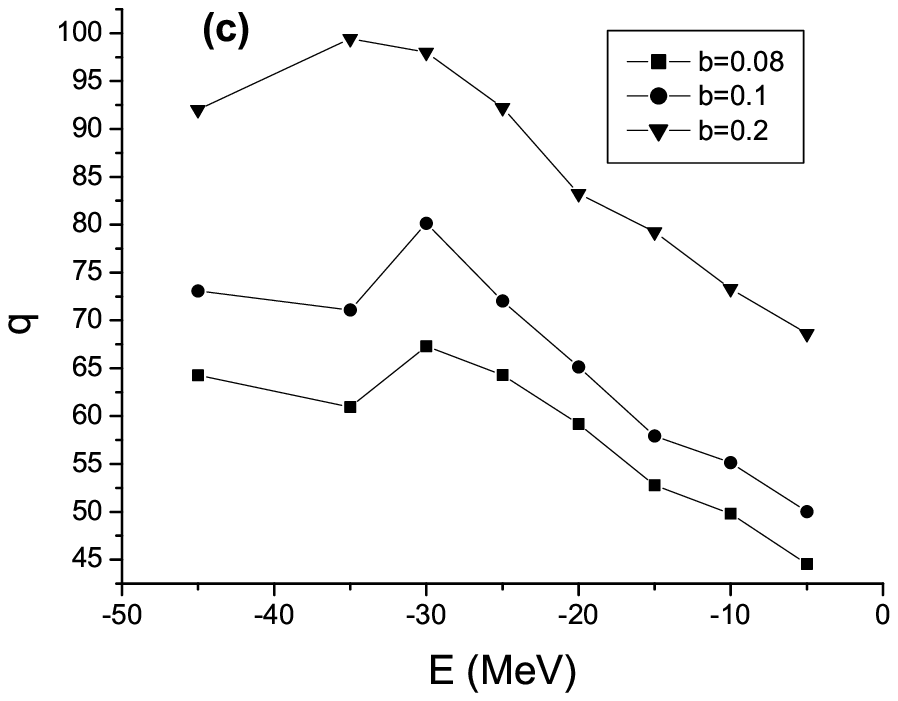}
\caption{(a) $<\lambda_1>$, (b) $T_{1/2}$ and (c) $q$ as a function of the total energy for $a=0.05 fm$ and $b=0.08$, $b=0.1$ and $b=0.2$. The other parameters are the same as those of Fig. 1.}
\end{center}
\end{figure}

Increasing $b$ further, will lead to a destruction of all KAM tori and thus to the usual behaviour of the relaxation time: $<\lambda_1>$ is an increasing function of the energy, q is also increasing due to the decrease of the area of the regular island around the $1:1$ resonance and $T_{1/2}$ is decreasing. As $b$ increases further ($b=3$ and above), although the portion of the chaotic phase space grows, some stickiness effects may occur which also leads to an increase of relaxation time \cite{VAR00}. 

Increasing $a$ leads to a decrease of the maximum value of the frequency of the Woods-Saxon potential and to a shift of its location towards lower energies. In Fig. 8, $(r,E_p)$ surfaces of section for $b=0.2$ and four values of $a$, namely $0.05 fm$, $0.5 fm$, $1.5 fm$ and $3.1 fm$, are shown. These correspond to the three different qualitative forms of the $\omega_p(E_p)$ curve (see Fig. 5). For $a=0.05 fm$ and $a=0.5 fm$ the curve has a maximum and two intersection points with the corresponding curve of the harmonic oscillator. The leftmost of these points corresponds to the  $1:1$ resonance which is located around $E_p=-30 MeV$ and can be seen in Figs. 8 (a) and (b). For $a=1.5 fm$, $\omega_p(E_p)$ has a maximum but no intersection points with the corresponding curve of the harmonic oscillator. In this case, as can be seen from Fig. 8 (b), chaos emerges first in the region of high particle energies. This is due to the fact that the  $\omega_p(E_p)$ curve is not symmetric with respect to its maximum: its slope is larger for energies above the maximum. Therefore, the density of the higher-order resonances ($1:2$, $1:3$, $\ldots$) is larger for these energies, and these resonances overlap, first creating a chaotic region at large energies. For $a=3.1$, $\omega_p(E_p)$ is decreasing and has one intersection point with the corresponding curve of the harmonic oscillator. In this case, the $1:2$, $1:3$, $\ldots$ resonances appear at higher energies and they become denser near the edge of the potential well, where the slope of the $\omega_p(E_p)$ curve becomes large.
\begin{figure}
\begin{center}

\caption{$(r,E_p)$ Poincar\'e sections for $b=0.2$ and (a) $a=0.05 fm$, (b) $a=0.5 fm$, (c) $a=1.5 fm$ and (d) $a=3.1 fm$. The other parameters are the same as those of Fig. 1.}
\end{center}
\end{figure}

The effect where Lyapunov exponent and relaxation time both increase can also be observed in simpler systems, provided that the dispersion relation $\omega(E)$ of the non-harmonic oscillator has an extended increasing part. Examples of such systems are oscillators with potentials $\sim x^{2n}$ ($n=2,3,\ldots$) coupled with harmonic oscillators. However, the generality of this effect is limited by the form of the coupling term. Actually it has been found that coupling terms exhibiting strongly chaotic behaviour (such as $\sim x^2y^2$), destroy the above mentioned effect and the usual behaviour of relaxation times is recovered.  

\section{Conclusions}
Motivated by questions of the damping mechanisms of collective motion in nuclei we have studied the behaviour of the relaxation time of an ensemble of chaotic orbits as a function of the total energy in a system consisting of two oscillators, one of them being harmonic. It has been found that, in a rather broad range of the parameters, the relaxation time can increase or remain constant although the Lyapunov exponent increases. We have attributed this behaviour to the appearance of KAM tori, which limit the phase space available to chaotic orbits. For this behaviour to occur, the dispersion relation of the non-harmonic oscillator must either be strictly increasing or have a maximum. The dependence of this behaviour on the parameters, as well as on the coupling has been investigated. In addition, it has been found that this behaviour can occur in very simple systems, provided that the coupling is not strongly chaotic. The model used in this work can be extended to study the direct decay of the isoscalar giant monopole resonances, i.e. the decay where nucleons can escape from the nucleus. A systematic study of this decay in several nuclei is presently under investigation \cite{PAP05}.

\begin{acknowledgments}
P.P. would like to thank the Greek Scholarships Foundation (IKY) for financial support. 
\end{acknowledgments}


\begin{thebibliography}{99}
\bibitem{LIC83} A.J. Lichtenberg and M. A. Lieberman, \emph{Regular and
    Stochastic Motion} (Springer-Verlag, New York, 1983).

\bibitem{TAB89} M. Tabor, \emph{Chaos and Integrability in Nonlinear
Dynamics}
  (John Wiley \& sons, 1989).

\bibitem{KAN94} H. Kandrup and M.E. Mahon, Phys. Rev. E {\bf 49}, 3735 (1994).

\bibitem{KANa04} H. Kandrup and S.J. Novotny, Cel. Mech. and Dyn. Astron. {\bf 88} 1 (2004).

\bibitem{KAN89} H. Kandrup, Phys. Lett. A {\bf 140}, 97 (1989).

\bibitem{KAN93} H. Kandrup, M.E. Mahon and H. Smit, Astron. Astrophys. {\bf 271} 440 (1993).

\bibitem{BER94} G.F. Bertch, R.A. Broglia,{\it Oscillations in finite quantum systems}, Cambridge University Press, Cambridge, England (1994).

\bibitem{BUR95} G.F. Burgio, M. Baldo, A. Rapisarda and P. Schuck, Phys. Rev. C {\bf 52}, 2475 (1995).

\bibitem{TSU96} T. Tsuchiya, N. Gouda and T. Konishi, Phys. Rev. E {\bf 53} 2210 (1996).

\bibitem{KAN97} H. Kandrup, Astrophys. J. {\bf 480} 155 (1997).

\bibitem{BAL98} M. Baldo, G.F. Burgio, A. Rapisarda and P. Schuck, Phys. Rev. C {\bf 58}, 2821 (1998).

\bibitem{LEP98} S. Lepri, Phys. Rev. E {\bf 58} 7165 (1998).

\bibitem{TSU00} T. Tsuchiya, N. Gouda, Phys. Rev. E {\bf 61} 948 (2000).

\bibitem{KAN03} H. Kandrup, I.M. Vass and I.V. Sideris, Mon. Not. R. Astron. Soc. {\bf 341} 927 (2003).

\bibitem{KAN04} H. Kandrup and C. Siopis, Mon. Not. R. Astron. Soc. {\bf 374} 957 (2004).

\bibitem{HAR04} M.N. Harakeh, Nucl. Phys. A {\bf 731} 411 (2004); A. Richter, {\it ibid} 59.

\bibitem{NEU01} P. von Neumann-Cosel, Nucl. Phys. A {\bf 687} 131c (2001); P.F. Bortignon, {\it ibid} 892c.

\bibitem{BLA80} J.P. Blaizot, Phys. Rep. {\bf 64} 171 (1980); G.F. Bertsch, P.F. Bortignon and R.A. Broglia, Rev. Mod. Phys. {\bf 55} 287 (1983).

\bibitem{BLO78} J. Blocki et al, Ann. Phys. {\bf 113} 330 (1978).

\bibitem{YOU99} D. Youngblood, H.L. Clark and Y.-W. Lui, Phys. Rev. Lett. {\bf 82} 691 (1999)

\bibitem{DRO95} S. Drozdz, S. Nishizaki, J. Wambach and J. Speth, Phys. Rev. Lett. {\bf 74} 1075 (1995).

\bibitem{VRE99} D. Vretenar, N. Paar, P. Ring, G.A. Lalazissis, Phys. Rev. E {\bf 60} 308 (1999).

\bibitem{PAP05} P.K. Papachristou, E. Mavrommatis, V. Constantoudis, F.K. Diakonos and J. Wambach, in preparation.

\bibitem{PAP00} T. Papenbrock, Phys. Rev. C {\bf 61} 034602 (2000).

\bibitem{SOS03} S.M. Soskin, R. Mannella, P.V.E. McClintock, Phys. Rep. {\bf 373} 247 (2003).

\bibitem{CON87} G. Contopoulos and C. Polymilis, Physics D {\bf 24} 328 (1987).

\bibitem{VAR00} K. Tsiganis, A. Anastasiadis and H. Varvoglis, Chaos, Solitons \& Fractals, {\bf 11} 2281 (2000).

\bibitem{BRA88} A. Bracco et al, Phys. Rev. Lett. {\bf 60} 2603 (1988); S. Brandenburg et al, Phys. Rev. C {\bf 39}, 2448 (1989).

\end{thebibliography}
\end{document}